\documentclass[prl,amsmath,aps,superscriptaddress,twocolumn]{revtex4-2}
\usepackage{amsfonts}
\usepackage{graphicx,multirow,xcolor}
\usepackage[colorlinks]{hyperref}
\hypersetup{
	colorlinks = true,
	urlcolor = {blue},
	citecolor = {blue},
	linkcolor= {red}
}

\newcommand{\ket}[1]{\left|{#1}\right\rangle}
\newcommand{\bra}[1]{\left\langle{#1}\right|}

\newcommand{\mV}{{\mathcal V}}

\newcommand{\mR}{{\mathcal R}}

\begin{document}

\title{Test of Genuine Multipartite Nonlocality}
\author{Ya-Li Mao}
\altaffiliation{These authors contributed equally to this work.}
\affiliation{Shenzhen Institute for Quantum Science and Engineering and Department of Physics, Southern University of Science and Technology, Shenzhen, 518055, China}
\affiliation{Guangdong Provincial Key Laboratory of Quantum Science and Engineering, Southern University of Science and Technology, Shenzhen, 518055, China}

\author{Zheng-Da Li}
\altaffiliation{These authors contributed equally to this work.}
\affiliation{Shenzhen Institute for Quantum Science and Engineering and Department of Physics, Southern University of Science and Technology, Shenzhen, 518055, China}
\affiliation{Guangdong Provincial Key Laboratory of Quantum Science and Engineering, Southern University of Science and Technology, Shenzhen, 518055, China}

\author{Sixia Yu}
\email{yusixia@ustc.edu.cn}
\affiliation{Hefei National Laboratory for Physical Sciences at Microscale and Department of Modern Physics, University of Science and Technology of China, Hefei,
	Anhui 230026, China}

\author{Jingyun Fan}
\email{fanjy@sustech.edu.cn}
\affiliation{Shenzhen Institute for Quantum Science and Engineering and Department of Physics, Southern University of Science and Technology, Shenzhen, 518055, China}
\affiliation{Guangdong Provincial Key Laboratory of Quantum Science and Engineering, Southern University of Science and Technology, Shenzhen, 518055, China}
\begin{abstract}
	
	While Bell nonlocality of a bipartite system is counter-intuitive, multipartite nonlocality in our many-body world turns out to be even more so. Recent theoretical study reveals in a theory-agnostic manner that genuine multipartite nonlocal correlations cannot be explained by any causal theory involving fewer-partite nonclassical resources and global shared randomness. Here we provide a Bell-type inequality as a test for genuine multipartite nonlocality in network  by exploiting a matrix representation of the causal structure of a multipartite system. We further present experimental demonstrations that both four-photon GHZ state and  generalized four-photon GHZ state significantly violate the inequality, i.e., the observed four-partite correlations resist explanations involving three-way nonlocal resources subject to local operations and common shared randomenss, hence confirming that nature is boundless multipartite nonlocal.
	
\end{abstract}
\maketitle

{\it Introduction --- } Nature allows nonlocal correlations between spacelike separated parties which cannot be explained by classical causal models. Nonlocality has been firmly established via the violation of the celebrated Bell Inequality~\cite{bell,chsh, bellnon} in a number of experiments with bipartite quantum systems~\cite{exp1,exp2,cexp2,cexp3,cexp4,cexp5,cexp6} and has led to critical applications in quantum information science such as quantum teleportation~\cite{tel0}, quantum key distribution \cite{qkd1,qkd3,qkd4}, and quantum randomness~\cite{ran1,ran2,ran5}. 
Going beyond, understanding nonlocality of a system with three or more parties is an intriguing question, which may potentially impact a broad range of applications such as multipartite cryptography~\cite{mcrp}, quantum computation~\cite{qc1,qc2,qc3}, correlating particles which never interacted~\cite{sw1,sw2}, many-body physics~\cite{cmp1,cmp2,cmp3,cmp4}, and quantum networks which have advanced signifiantly in the past a few years~\cite{qn1,qn2,qn3,qn4,qn5,qn6,qn7,qn8,qn9,qn10,qn11,qn12,qn13,qn14,qn15,qn16,netwm3,netwm5,qn17}, besides deepening our understanding of nonlocality.  

Multipartite systems have much richer correlation structures comparing with bipartite systems. According to Svetlichny's  proposal of genuine multipartite nonlocality~\cite{svet} restricted by non-signaling conditions~\cite{op1,op2}, it is possible to construct genuine multipartite correlations with bipartite resources \cite{intrinsic}. Actually Svetlichny's original proposal provided a device-independent witness of genuine multipartite entanglement~\cite{royl,roya}, in which he adopted the framework of local operation and classical communications (LOCC). However, for spacelike separated parties in multipartite Bell scenarios classical communications are not at their disposal, which enforces no-signaling condition. Furthermore, it is a realistic scenario that all parties may have global access to common shared randomess. Hence, it is of particular interest to ask whether there are  multipartite nonlocal correlations that cannot be explained with bipartite and any other fewer-partite nonlocal resources which are subject to local operations (without classical communications) and shared randomness (LOSR)~\cite{losr0,losr00,losr1,losr2,losr3}. This question led to the latest theoretical advances of genuine LOSR network multipartite entanglement~\cite{ngme} and genuine LOSR network multipartite nonlocality~\cite{chao,royl,roya,bier}. 

In~\cite{royl,roya}, Coiteux-Roy, Wolfe, and Renou defined genuine LOSR multipartite nonlocality, referred to as  genuine multipartite nonlocality in network here, to be those correlations that cannot be simulated with local composition of any $k-$partite (with $k\in\{1,2,...,N-1\}$) resources with access to common shared randomness for a network with $N$ parties. 
From the theory-agnostic perspective, they considered any plausible causal theory including classical theory, quantum theory and hypothetical generalized probabilistic theory (GPT) that is compatible with device replication. Exploiting the inflation techniques~\cite{infl1,infl2,infl3}, they designed a device-independent Bell-type inequality for genuine LOSR multipartite nonlocality~\cite{royl,roya} and proved that $N-$partite GHZ state can violate their inequality and thus is genuine LOSR multipartite nonlocal for any finite $N$. This line of research promotes our understanding of nonlocality by revealing that nature is boundlessly nonlocal and in the meantime showcases the usefulness of inflation technique. 

In this Letter, we first propose an algebraic  approach to inflation technique via matrix representation of the causal structure, i.e., party-resource relations of a general network with $N$ parties. This enables the construction of a new test of genuine LOSR multipartite nonlocality in terms of Bell-type inequality, in which each party performs two alternative dichotomic measurements. This test outperforms the one proposed in~\cite{royl} with an improved noise threshold which attains the optimal value found via linear programming in~\cite{roya} for tripartite GHZ state. Furthermore, we experimentally demonstrate that four-photon and three-photon GHZ states and the respective generalized GHZ states violate the inequalities. Finally we conclude with a discussion that a large family of quantum pure states can violate the Bell-type inequality besides quantum GHZ state and $W$ state~\cite{royl,roya}. 

\begin{figure}
	\centering
	\includegraphics[width=1\linewidth]{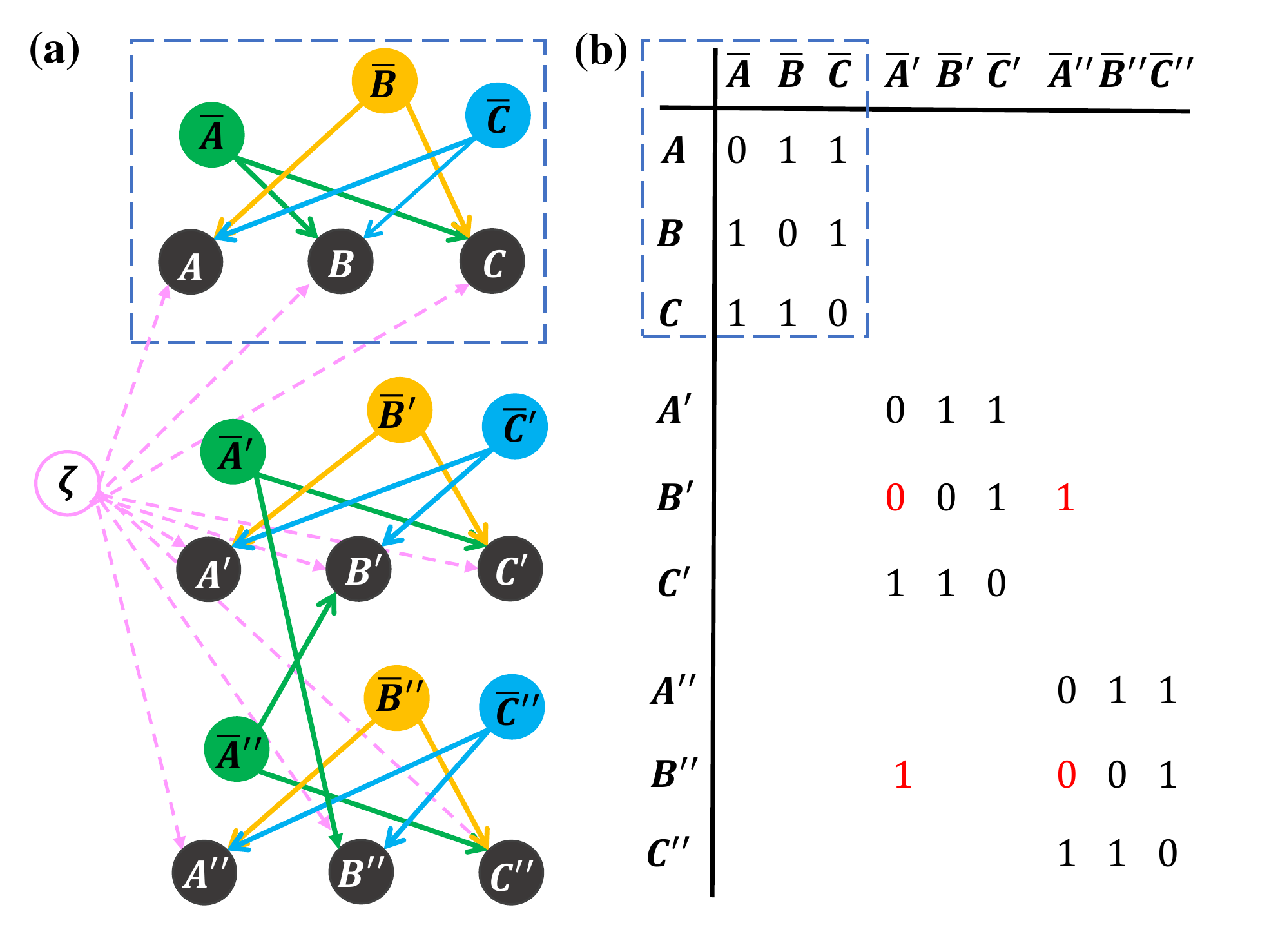}
	\caption{ A triangular network with parties $\mV=\{A,B,C\}$ and resources $\mR=\{\bar A,\bar B,\bar C\}$ (enclosed by dashed line) and its non-fan-out-inflation of order 3 (entire) in (a) with their respective matrix representations $\Gamma$ (enclosed by dashed line) and $\Gamma'$ in (b), where blank entries represent zero. All parties have access to common shared randomness ($\zeta$).
	}
	\label{FIG1}
\end{figure}

{\it Test in triangular network --- }
In the framework of LOSR, we consider first a network of three parties, labelled with $\mV=\{A,B,C\}$, with global access to common shared randomness $\zeta$ as shown in Fig. \ref{FIG1}(a). Each pair of parties share a two-way  resource, namely, $\omega_{AB} \equiv\bar C$, $\omega_{BC} \equiv\bar A $, and $\omega_{CA} \equiv\bar B$, which can be very general, such as classical, quantum, or no-signaling. As shown in Fig. \ref{FIG1}(b), the network can be faithfully represented by a $3\times3$ incidence matrix $\Gamma$ with elements $\Gamma_{ij}=1-\delta_{ij}$ for $i,j=0,1,2$. A matrix element $\Gamma_{ij}=1(0)$ indicates that the $i$-th party in the ordered set $\{A,B,C\}$ is (not) sharing the $j$-th resource in the ordered set $\mR=\{\bar A,\bar B,\bar C\}$. 
Following Ref.~\cite{royl, roya}, based on the assumption of device replication and causality, a non-fan-out inflation of the triangular network of order 3 is 
a network of nine parties $\{\mV,\mV',\mV''\}$ connected by resources $\{\mR,\mR',\mR''\}$. Its faithful incidence matrix representation $\Gamma'$ is presented in Fig. \ref{FIG1}(b), in which each row and column have exactly $N-1$ nonzero entries, one for each type of resources and parties, respectively. 

In the triangular network,  each party performs two alternative dichotomic measurements $\mV_{\bf x}=\{A_x,B_y,C_z\}$ (respectively with random and private inputs $x,y,z=0,1$) with outcomes ${\bf a}=\{a,b,c\}\in\{-1,1\}$, giving rise to a set of correlations 
$ P({\bf a}|\mV_{\bf x})$. 
In the inflated network, the measurements for the same type of parties are identical, e.g., measurements $A'_x$ and $A''_x$ performed by parties $A',A''$ are  the same as $A_x$, so do other parties, respectively. As a result, we obtain an inflated correlation 
$\mathcal Q_3({\bf aa'a''}|\mV_{\bf x}\mV'_{\bf x'}\mV''_{\bf x''})$ satisfying a set of compatibility rules. First, it is no-signaling for all parties. Second,
as a consequence of no-signaling and causality, as long as two subnetworks are isomorphic the  correlations among the corresponding parties are identical. In an inflated network, two subnetworks  are isomorphic if they are isomorphic under the dropping of the primes of the parties and resources \cite{roya}. For instance we have
$\langle A'B'\rangle_{\mathcal Q_3}=\langle AB\rangle_{\mathcal Q_3}$, $\langle A'C'\rangle_{\mathcal Q_3}=\langle AC\rangle_{\mathcal Q_3}$, and
$\langle ABC\rangle_{\mathcal Q_3}=\langle ABC\rangle_P.$
And lastly it is nonnegative, e.g., 
\begin{equation}
	\sum_{\stackrel{\alpha,\beta=\pm1}{y=0,1}}\mathcal Q_3((-1)^y\alpha\beta,-\beta,\alpha,\beta|A_1B_yC_1C_0')\ge0,
\end{equation}
from which it follows that, 
\begin{equation}
	2-\langle A_1(B_0-B_1)C_1+(B_0+B_1)C_0'\rangle_{\mathcal Q_3}\ge0. 
\end{equation}

To proceed we note that
\begin{eqnarray*}
	\langle B_yC_0'\rangle_{\mathcal Q_3}&=&\langle B_y' C_0'\rangle_{\mathcal Q_3}\\ &\ge& \langle B_y'A_0'\rangle_{\mathcal Q_3}+\langle A_0' C_0'\rangle_{\mathcal Q_3}-1\\&=&
	\langle B_yA_0+A_0 C_0\rangle_P-1.
\end{eqnarray*}
Here the first equality holds because of isomorphism $\{B,C'\}$ and $\{B',C'\}$ both of which do not share a common nonlocal resource and locally they have the same pattern of resources sharing, which is evident in matrix $\Gamma'$ (see Lemma later in the text). The inequality comes from positivity $\sum_{\pm}\mathcal Q_3(\mp,\pm,\pm| A_0' B_y' C_0')\ge0. $
The last equality is due to isomorphisms such as $\{A'B'\}$ with $\{AB\}$ and compatibility.
Finally we obtain the following Bell-type inequality as a test of LOSR genuine three-partitie nonlocality in a triangular network,
\begin{equation}\label{3}
	\langle A_0(B_0+B_1)+A_1(B_0-B_1)C_1+2A_0C_0\rangle_P\le 4.
\end{equation}

\begin{figure*}
	\centering
	\includegraphics[width=\linewidth]{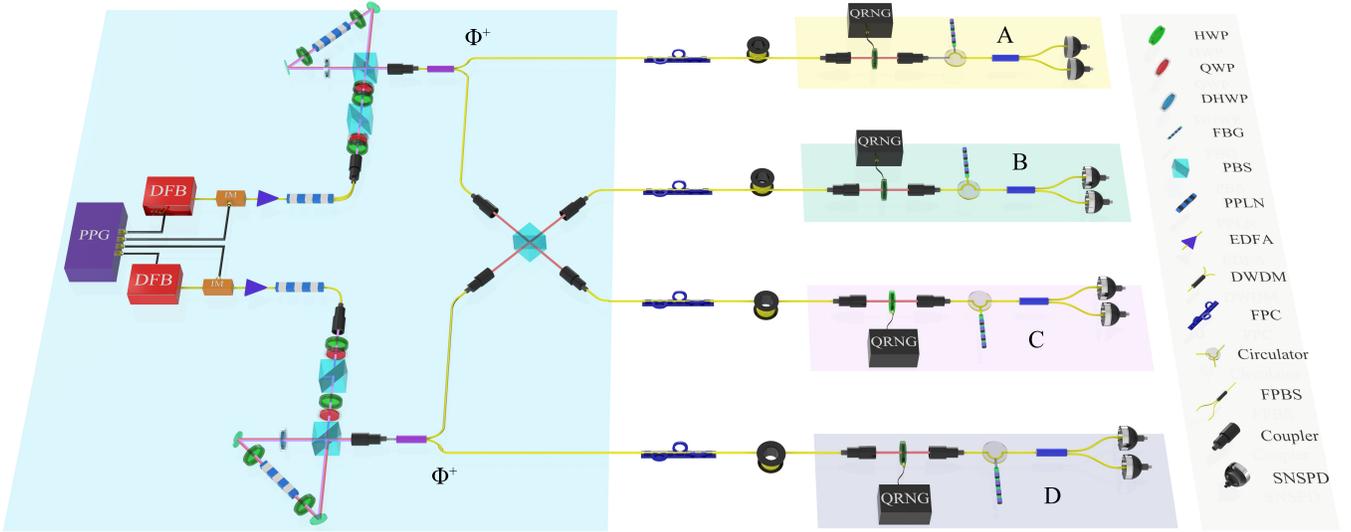}
	\caption{Schematic of the experiment. We prepare two EPR sources by driving two SPDC processes in parallel (see text for details). In each SPDC process, we inject a laser pulse at 779 nm into a PPLN crystal in a Sagnac interferometer~\cite{Sun2019}, which probabilistically emits a pair of polarization-entangled photons in EPR state $\ket{\Phi^+}$ at paired wavelengths 1556 nm (signal) and 1560 nm (idler). We interfere signal photons from the two sources on a PBS and obtain four-photon GHZ state sharing between parties A, B, C, and D after post-selection. A quantum random number generator (QRNG) is used to instruct each party to perform two alternative dichotomic measurements to the photon at his disposal. DHWP: Dual-wavelength HWP for 1560 nm and 779 nm; DWDM: Dense wavelength division multiplexing; FPBS: Fiber PBS; FPC: fiber polarization controller; SNSPD: superconducting-nanowire-single-photon-detector.
	}
	\label{FIG2}
\end{figure*}

Some remarks are in order. First, it is straightforward to show that substituting GHZ state $|\text{GHZ}_3\rangle=\left|000\rangle+|111\rangle\right/\sqrt{2}$ under local measurements $A_0=C_0=Z$, $A_1=C_1=X$, and $B_{y}=(Z+(-1)^yX)/\sqrt2$ to the left hand side of Inequality~(\ref{3}) yields $2+2\sqrt2$, hence violating the inequality. (Note that we use $\ket{0}(\ket{1})$ to denote photon in horizontal (vertical) polarization state $\ket{H}(\ket{V})$ and $X,Y,Z$ are Pauli matrices.)

Second, each party performs two alternative dichotomic measurements in our test, in contrast to the original proposals in which one party performs three alternative dichotomic measurements~\cite{royl,roya}. This enables us to find the device independent maximal violation to the inequality, hence providing a device independent detection of genuine multipartite nonlocality. It turns out that the maximum is attained at projective measurements performed on a 3-qubit pure state~\cite{qubit}. Thus, for quantum theory ($Q$), we have (see Supplemental Material)
$\langle A_0(B_0+B_1)+A_1(B_0-B_1)C_1+2A_0C_0 \rangle_P\stackrel{Q} {\le} 2+2\sqrt2.$ 
Clearly GHZ state provides the maximal violation. Interestingly the algebraic upper bound of this inequality is 6, which is attained by the extremal Box 8 as documented in Ref.\cite{box}.

Third, by symmetry we can obtain other Bell-type inequalities for genuine multipartite nonlocality in network by exchanging some parties, e.g., $A$ and $C$.

{\it General network --- }
As isomorphic subnetworks give rise to identical correlations, it is critical to identify isomorphic subnetworks in designing tests for genuine LOSR multipartite nonlocality in network. The following Lemma provides a criterion for isomorphism among two-party subnetworks exploiting the matrix representation for inflated network.

{\bf Lemma } Consider an inflated  network of order $k$ with $kN$ parties  $\{ N_\mu\}_{\mu=0}^{kN-1}$ specified by $kN\times kN$ incidence matrix $\Gamma$. The same type of parties and resources are labeled with indices having the same remainder modular $N$. Two subnetworks $\{N_\mu, N_\nu\}$ and $\{ N_{\mu'}, N_{\nu'}\}$ are isomorphic iff  $\vec\gamma_{\mu\nu}=\vec\gamma_{\mu'\nu'}$ where the $N$-dimensional vector $\vec\gamma_{\mu\nu}$ is defined by components
\begin{equation}
	[\vec\gamma_{\mu\nu}]_s:=\sum_{t\equiv s\ ({\rm mod}\, N)}\Gamma_{\mu t}\oplus_2\Gamma_{\nu t}
\end{equation}
for $0\le s\le N-1$ where $\oplus_2$ denotes addition modular 2.

The proof is straightforward by noting that each component of vector $\vec\gamma_{\mu\nu}$ can assume only 3 possible values $\{0,1,2\}$. It is zero iff they share the corresponding resource and 1 iff the corresponding resource is  $\bar N_\mu$ and $\bar N_\nu$ and $2$ iff the corresponding resources are not shared.
For examples, in the inflated network in Fig. \ref{FIG1}, we have isomorphic subnetworks $\{B,C'\} \sim \{B',C'\}$ and $\{A,B\}\sim \{A',B'\}$, which is evident given $\vec \gamma_{BC'}=\vec\gamma_{B'C'}=(2,1,1)$ and $\vec\gamma_{AB}=\vec\gamma_{A'B'}=(1,1,0)$. Here we use $\sim$ to denote that the two subnetworks have the same causal structures. Similarly we have $\{A,C\}\sim \{A',C'\}$, while subnetworks $\{B,C\}$ and $ \{B',C'\}$ are not isomorphic.

Equipped with this Lemma, we can extend the Bell-type inequality for triangular network to a general network with $N$ paties (see Supplemental Material).

{\bf Theorem } For a general network with $N$ parties, labelled with $\{A,B,C,D,\ldots,W\}$, with each group of $N-1$ parties sharing a nonlocal resource in addition to global randomness, the correlation produced by two local dichotomic measurements on each party satisfies the following Bell-type inequality 
\begin{widetext}
	\begin{eqnarray}
		S_N:= \langle A_0(B_0+B_1)+2(A_0C_0+C_0D_0+\ldots+V_0W_0)
		+A_1(B_0-B_1)C_1D_1\ldots W_1\rangle_P &\stackrel{LOSR}\le& 2(N-1), \label{Nbit}\\
		&\stackrel{Q}{\le}& 2\sqrt2+2(N-2).
	\end{eqnarray}
\end{widetext}
The LOSR bound is maximally violated by $N$-qubit GHZ state with a white-noise threshold 
$\eta_N=\frac{N-1}{N-2+\sqrt 2}$, which improves over previous results~\cite{royl} (see Supplemental Material). For example, we obtain $\eta_3=0.83$ which is smaller than 0.93 in~\cite{royl} and coincides with the optimal threshold found via linear programming in~\cite{roya}, hence confirming that our analytical results are optimal. We shall present below experimental verification of our results in the cases of $N=3,4$.

\begin{figure}[htbp!]
	\centering
	\includegraphics[width=\linewidth]{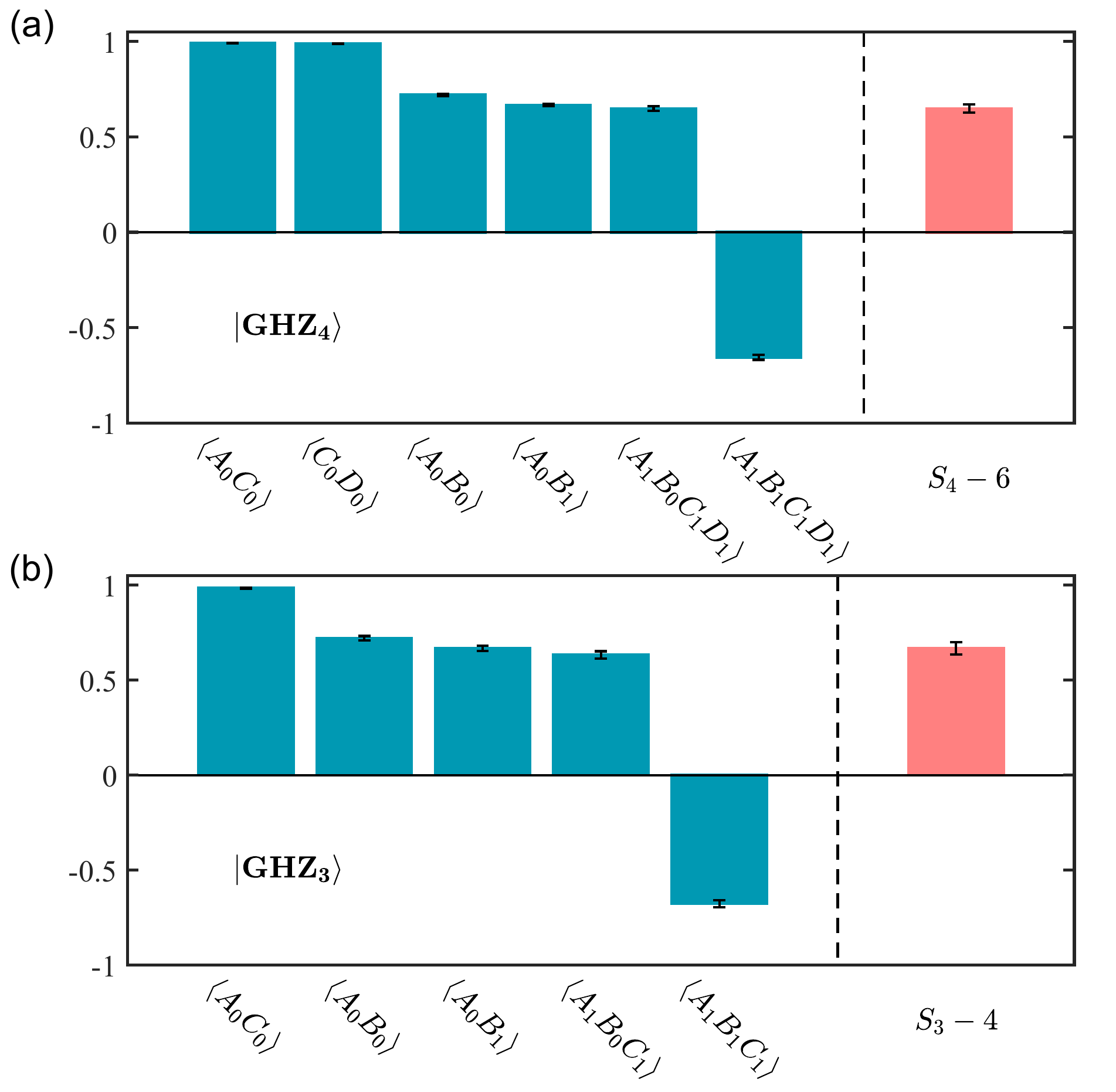}
	\caption{Experimental measurements of correlation function $S_N$ (left panel) and violations of Inequality (\ref{Nbit}), $S_N-2(N-1)$, (right panel) with GHZ states $N=4$ in (a) and $N=3$ in (b). Error bars represent one standard deviations in experiments.}
	\label{FIG3}
\end{figure}

{\it Experiments --- }A schematic of implementing a quantum network distributing four-photon GHZ state to four parties Alice (A), Bob (B), Charlie (C) and David (D) is depicted in Fig. \ref{FIG2}. We first prepare two EPR sources. We use a pulse pattern generator (PPG) to send out trigger pulses at a rate of 250 MHz. In each source, the trigger pulse signals a distributed feedback (DFB) laser to emit a laser pulse at  $\lambda=1558$ nm. We shorten the pulse width from 2 ns to 90 ps with an intensity modulator (IM). After passing through an erbium-doped-fiber-amplifier (EDFA), a periodically poled MgO doped Lithium Niobate (PPLN) waveguide to double the frequency, and a dense wavelength division multiplex (DWDM) filter to remove the residual pump light, we use the produced pulse at $\lambda_p=779$ nm to drive a Type-0 spontaneous parametric downconversion (SPDC) process in a piece of PPLN crystal in a Sagnac interferometer, which emits probabilistically a pair of photons in EPR state $\ket{\Phi^+}=(\ket{HH}+\ket{VV})/\sqrt{2}$~\cite{Sun2019} at the phase-matched wavelength 1556~nm (signal) an 1560~nm (idler). 
Interfering signal photons from the two EPR sources on a polarizing beamsplitter (PBS), we create four-photon GHZ state,  $\ket{\text{GHZ}_4}=(\ket{HHHH}+\ket{VVVV})/\sqrt{2}$, after post-selection~\cite{Pan2001PRL}. We pass photons through fiber Bragg gratings (FBGs) with bandwidths of 3.3 GHz before entering single-photon detectors to suppress the spectral distinguishability between photons from different EPR sources and keep the photon-pair production rate of each EPR source at 0.0025 per trigger to strongly mitigate the multi-photon effect. Quantum tomography measurements indicate that the state fidelity is greater than $0.99$ for the two-photon states produced at EPR sources with respect to the ideal Bell state $|\Phi^+\rangle$ and is $0.9740\pm0.0043$ for the produced four-photon state with respect to the ideal state $\ket{\text{GHZ}_4}$, respectively. 

\begin{figure}[htbp!]
	\centering
	\includegraphics[width=\linewidth]{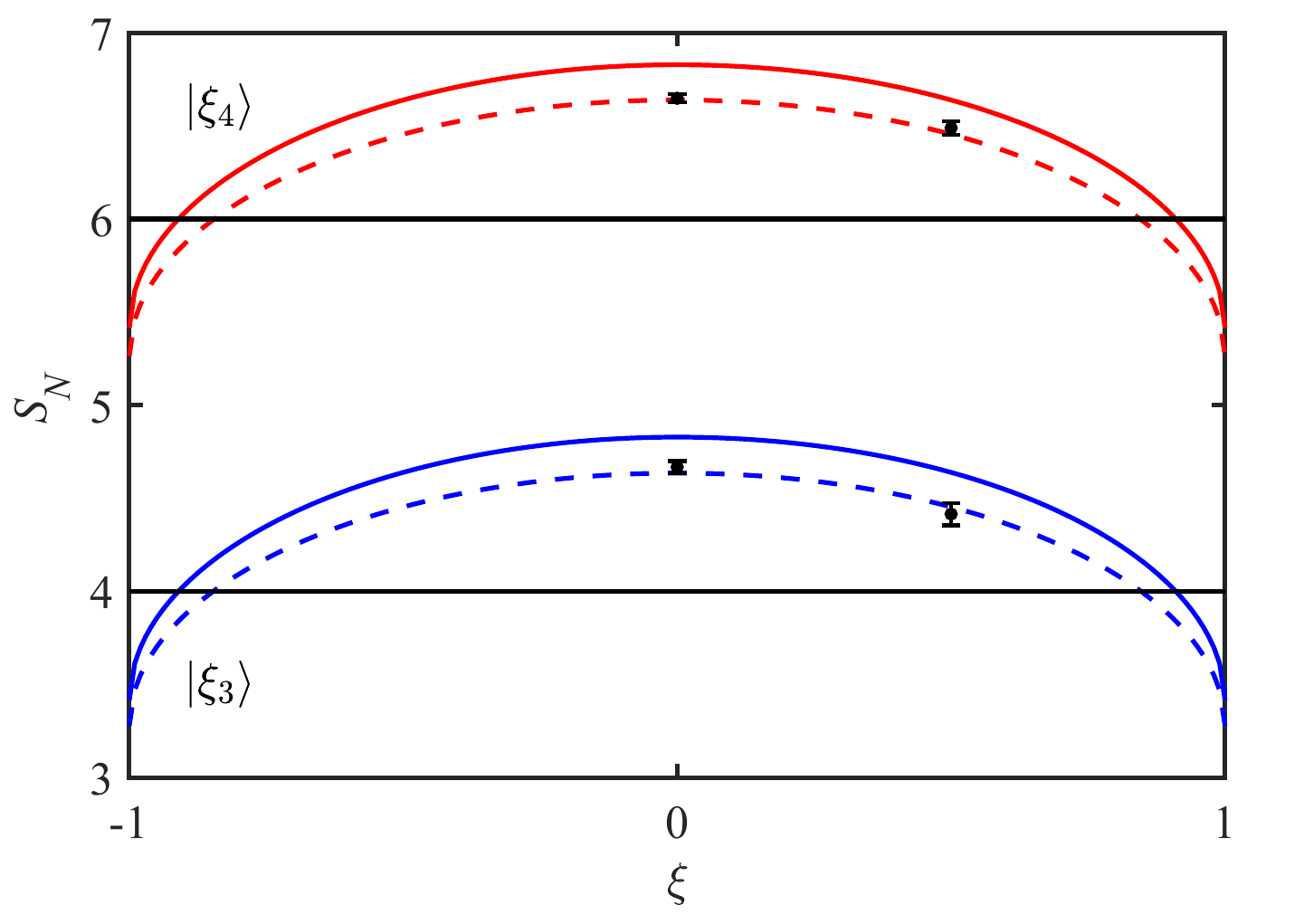}
	\caption{Correlation function $S_N$ for generalized GHZ states $\ket{\xi_3}$ (blue, bottom) and $\ket{\xi_4}$ (red, upper) with the respective LOSR bounds (horizontal lines). Smooth lines: ideal theory, dashed lines: theory considering white noise ($\eta_N$) in the experiment, filled dots: $S_4=6.6484\pm0.0209$ ($6.4890\pm 0.0375$) and $S_3=4.6674\pm 0.0323$ ($4.4153\pm0.0603$) when $\xi=0$ ($0.5$), measured with one standard deviations in experiment.
	}
	\label{FIG4}
\end{figure}

We install a quantum random number generator at each party~\cite{ZDLi}, which privately and randomly feeds a 2-bit random number ($x_\alpha\in\{0,1\}$) to the party to switch between measurement bases $Z$ ($x_\alpha=0$) and $X$ ($x_\alpha=1$) for Alice, Charlie, and David, and between $(Z+X)$ ($x_\alpha=0$) and $(Z-X)$ ($x_\alpha=1$) for Bob to perform measurement to her/his share of photon, where the Pauli matrices $X$, $Z$, and $Z \pm X$ are implemented by a half-wave plate (HWP) at each party, resepectively. The generation of random numbers is synchronized to the PPG. We switch the measurement settings every 30 seconds, reserving the frst 10 seconds to reset the measurement settings, including quantum random number generation and waveplate rotation, and the remaining 20 seconds for data collection. 
We collect $33252$ four-photon coincidence events over 16 measurement setting combinations in $14741$ switching cycles. We compute the correlation function $S_4=6.6484\pm 0.0209$ which surpasses the LOSR bound 6 by more than 30 standard deviations as shown in Fig. \ref{FIG3}(a), i.e., the observed correlation cannot be reproduced by involving any three-way nonlocal resources with local operations and unlimited shared randomness. Hence the observed correlation is geuninely LOSR four-partite nonlocal.  

Each of David's succesful detection of a photon probabilitically heralds the presence of a three-photon GHZ state shared between Alice, Charlie, and Bob. We show in Fig.~\ref{FIG3}(b) the correlation function $S_3=4.6674\pm0.0323$ surpasses the LOSR bound 4 by more than 20 standard deviations, i.e., the observed correlation cannot be reproduced by involving any two-way nonlocal resources with local operations and unlimited shared randomness.

{\it Discussions and conclusions --- }Besides GHZ state and $W$ state~\cite{royl,roya}, we now show that a large family of pure states can violate Inequality (\ref{Nbit}). 
First, it is straightforward to show that the generalized $N$-qubit GHZ state $\ket{\xi_N}= \sqrt{\frac{1+\xi}2}|0\rangle^{\otimes N}+\sqrt{\frac{1-\xi}2}|1\rangle^{\otimes N}$ violates Inequality (\ref{Nbit}) with the same measurement settings as those for GHZ state, whenever $|\xi|<\xi_c$ with $|\xi|\in[-1,1]$ and $\xi_c=0.91$ independent of $N$. We present in Fig.~\ref{FIG4} experimental demonstrations for $N=3$ and $N=4$ with $\xi=0.5$ along with theoretical predictions. The results uphold a good agreement. 

Furthermore, under local unitaries, the most general 3-qubit pure state can be cast into the canonical form
$$|\Psi_3\rangle=h_0|000\rangle+h_1e^{i\phi}|100\rangle+h_2|101\rangle+h_3|110\rangle+h_4|111\rangle$$
with $h_i\ge0$ and $\sum_ih_i^2=1$ and $\phi\in[0,\pi]$. Performing the same measurements to this state as that for GHZ state shall violate Inequality (\ref{3}) if the state satisfies the condition  $\sqrt2((h_0+h_4)^2+2h_3^2+h_0^2+h_4^2-1)+4(h_0^2+h_4^2+h_2^2)-2>4$. 

It is reasonable for one to anticipate that the study of multipartite nonlocality may be as fruitful as that of Bell nonlocality~\cite{bellnon}. Hence, it will be interesting to explore more states and new approaches suitable for the test of genuine LOSR multipartite nonlocality, for example, Hardy-type of nonlocality tests, which have been used to detect genuine multipartite nonlocality in Svetlichny's original definition with no-signaling restrictions~\cite{y1,y2}. The matrix representation of the causal relation of networks introduced in this work may provide a convenient tool in these explorations. 

{\it Acknowledgement ---} This work is supported by the Key-Area Research and Development Program of Guangdong Province Grant No.2020B0303010001, Grant No.2019ZT08X324, Guangdong Provincial Key Laboratory Grant No.2019B121203002, and the National Natural Science Foundation of China Grants No.12004207 and No.12005090.

\textit{Note added.---}While finishing this manuscript, we became aware of a related work by Huan Cao et al~\cite{Huan2022}.

\clearpage

\onecolumngrid
\subsection*{\textbf{\large Supplementary Materials for Test of Genuine Multipartite Nonlocality}}

\section{Proof of Theorem in the main text}

Consider a network with $N$ nodes labeled with $$\mathcal V=\{A,B,C,\ldots,V,W\}$$
that are connected by $(N-1)$-shared nonlocal resources
$$\mathcal R=\{\bar A,\bar B,\bar C,\ldots,\bar V,\bar W\}$$ 
where, e.g., $\bar A\equiv\omega_{BCD\ldots W}$, denotes the $(N-1)$-shared nonlocal resource  not involving $A$. The original network connected by these nonlocal resources can  be specified by the following incidence matrix
\begin{equation}\begin{array}{c|cccccc}&\bar B&\bar A&\bar C&\ldots &\bar V&\bar W\\\hline
		B&0&1&1&\cdots&1&1\\
		A&1&0&1&\cdots&1&1\\
		C&1&1&0&\cdots&1&1\\
		\vdots&\vdots&\vdots&\vdots&\ddots&\vdots&\vdots\\
		V&1&1&1&\ldots&0&1
		\\
		W&1&1&1&\ldots&1&0
\end{array}\end{equation}
where we have exchanged nodes $B$ and $A$ and resources $\bar A,\bar B$ for later use.

\begin{table}
	[h]
	$$\begin{array}{c|c@{\hskip3.5pt}c@{\hskip3.5pt}c@{\hskip3.5pt}c@{\hskip3.5pt}c@{\hskip3.5pt}c@{\hskip3.5pt}c|c@{\hskip2pt}c@{\hskip 2pt}c@{\hskip 2pt}c@{\hskip 2pt}c@{\hskip 2pt}c@{\hskip 2pt}c}&\bar B'&\bar A'&\bar C'&\bar D'&\ldots &\bar V'&\bar W'&\bar B''&\bar A''&\bar C''&\bar D''&\ldots&\bar V''&\bar W''\\\hline
		B'&0&\textcolor{red}0&\textcolor{red}0&\textcolor{red}0&\ldots&\textcolor{red}0&1&&\textcolor{red}1&\textcolor{red}1&\textcolor{red}1&\ldots&\textcolor{red}1 \\
		A'&1&0&\textcolor{red}0&\textcolor{red}0&\ldots&\textcolor{red}0&1 &&&\textcolor{red}1&\textcolor{red}1&\ldots&\textcolor{red}1\\
		C'&1&1&0&\textcolor{red}0&\ldots&\textcolor{red}0&1&&&&\textcolor{red}1&\ldots&\textcolor{red}1\\
		D'&1&1&1&0&\ldots&\textcolor{red}0&1&&&&&\ldots&\textcolor{red}1\\
		\vdots&\vdots&\vdots&\vdots&\vdots&\ddots&\vdots&\vdots&&&&&&\vdots\\
		V'&1&1&1&1&\ldots&0&1&&&\\
		W'&1&1&1&1&\ldots&1&0&&&\\\hline
		B''&&\textcolor{red}1&\textcolor{red}1&\textcolor{red}1&\ldots&\textcolor{red}1&&0&\textcolor{red}0&\textcolor{red}0&\textcolor{red}0&\ldots&\textcolor{red}0&1 \\
		A''&&&\textcolor{red}1&\textcolor{red}1&\ldots&\textcolor{red}1&&1&0&\textcolor{red}0&\textcolor{red}0&\ldots&\textcolor{red}0&1 \\
		C''&&&&\textcolor{red}1&\ldots&\textcolor{red}1&&1&1&0&\textcolor{red}0&\ldots&\textcolor{red}0&1\\
		D''&&&&&\ldots&\textcolor{red}1&&1&1&1&0&\ldots&\textcolor{red}0&1\\
		\vdots&&&&&&\vdots&&\vdots&\vdots&\vdots&\vdots&\ddots&\vdots&\vdots \\
		V''&&&&&&&&1&1&1&1&\ldots&0&1\\
		W''&&&&&&&&1&1&1&1&\ldots&1&0
	\end{array}$$
	\caption{The incidence matrix of an inflation of order 2 (as a subnetwork of an inflation of order 3) of a network with $N$ nodes connected with $(N-1)$ nonlocal resources.  We have exchanged the position of nodes $A$ and $B$ and corresponding nonlocal resources so that all the elements in the upper triangle are red.}
\end{table}
Consider now an inflation $\mathcal Q_3$ of order 3  with $3N$ nodes $\{\mathcal V, {\mathcal V'},\mathcal V''\}$. On nodes in $\mathcal V$ we build an identical subnetwork to the original one, i.e., with incidence matrix given above. On nodes $\{{\mathcal V'},\mathcal V''\}$ we build a network with the  incidence matrix as specified in Table.I, which represents a legit inflation as each type of nodes, e.g., $\{A',A''\}$ are connected to once and only once the resources that involving the given node, e.g., $\{\bar B,\bar C,\ldots,\bar V,\bar W\}$  primed or doubly primed.  According to Lemma we have isomorphism $\{B,W'\}\sim \{B',W'\}$ as $\vec \gamma_{BW'}=\vec\gamma_{B'W'}=(1,2,2,\ldots,2,1)$ and isomorphisms $\{A,B\}\sim\{A',B'\}$ and $\{A C\}\sim\{A',C'\}$ and so on to $\{V,W\}\sim \{V',W'\}$ as we have, e.g., $\vec\gamma_{AC}=\vec\gamma_{A'C'}=(0,1,1,0,\dots,0)$.

Let each party, e.g. $A$, perform two alternative dichotomic measurements, e.g., $A_{0,1}$, giving rise to correlation $P({\bf a}|\mathcal V_{\bf x})$ where ${\bf a}=(a_A,a_B,\ldots,a_W)$ denotes the outcomes labeled with $\pm1$ and $\mathcal V_{\bf x}=\{A_x,B_y,\ldots,W_w\}$ denotes the measurement settings with $x,y\ldots,w\in\{0,1\}$. In the inflated network of order 3 specified above we have correlation
$\mathcal Q_3({\bf a}{\bf a}'{\bf a}''|\mathcal V_{\bf x}\mathcal V'_{\bf x'}\mathcal V''_{\bf x''})$, which satisfies a set of compatibility constraints arising from the assumptions of device replication and causality \cite{royl,roya}. First the measurements performed on each node of the same type are identical as a result of device replication. Second, nonsignaling conditions are imposed on all the nodes $\{\mathcal V,\mathcal V',\mathcal V''\}$.  Third, isomorphic subnetworks give rise to identical correlations as a result of causality. Therefore as subnetwork on $\mathcal V$ is identical to the original network, we have identical correlations $\mathcal Q_3({\bf a}|\mathcal V_{\bf x})=P({\bf a}|\mathcal V_{\bf x})$. Moreover  we also have identical correlations (in terms of expectation values) $\langle B_yW_0'\rangle_{\mathcal Q_3}=\langle B_y'W_0'\rangle_{\mathcal Q_3}$ and 
\begin{eqnarray*}
	\langle   A'  B'\rangle_{\mathcal Q_3}&=&\langle AB\rangle_{P}\\
	\langle   A'  C'\rangle_{\mathcal Q_3}&=&\langle AC\rangle_{P}\\
	\langle   C'  D'\rangle_{\mathcal Q_3}&=&\langle CD\rangle_{P}\\
	&\vdots&\\
	\langle   V'  W'\rangle_{\mathcal Q_3}&=&\langle VW\rangle_{P}
\end{eqnarray*}
for isomorphic subnetworks with two nodes.  Lastly, we have positivity, e.g.,
\begin{eqnarray*}
	\sum_{\stackrel{\alpha_B\alpha_C\ldots\alpha_V\alpha_W=\alpha}{y=0,1;\alpha,\beta=\pm1}}\mathcal Q_3((-1)^y\alpha\beta,\alpha_B,\alpha_C,\ldots,\alpha_V,\alpha_W,\beta|A_1B_yC_1\ldots V_1W_1  W_0')\ge0
\end{eqnarray*}
from which it follows that
\begin{eqnarray*}
	2-\langle A_1(B_0-B_1)C_1D_1\ldots W_1\rangle_{\mathcal Q_3}
	&\ge&\langle(B_0+B_1)  W_0'\rangle_{\mathcal Q_3}=
	\sum_{y=0}^1\langle   B_y ' W_0'\rangle_{\mathcal Q_3}\\
	&\ge&\sum_{y=0,1}\big(\langle  B_y ' A_0'+  A_0'  W_0'\rangle_{\mathcal Q_3}-1\big)\\
	&\ge&\sum_{y=0,1}\big(\langle  B_y'  A_0'+  A_0'  C_0'+  C_0'  W_0'\rangle_{\mathcal Q_3}-2\big)\ge \cdots\\
	&\ge&\sum_{y=0,1}\big(\langle  B_y'  A_0'+  A_0'  C_0'+  C_0'  D_0'+\ldots+  V_0'  W_0'\rangle_{\mathcal Q_3}-N+2\big)\\
	&=&\sum_{y=0,1}\big(\langle B_yA_0+A_0C_0+ C_0 D_0+\ldots+V_0W_0\rangle_P-N+2\big)
\end{eqnarray*}
proving the upper bound for LOSR. Here we have recursively made use of the fact that $\langle UV\rangle_{\mathcal Q}\ge \langle UW\rangle_{\mathcal Q} +\langle VW\rangle_{\mathcal Q}-1$ for a nonsignaling correlation ${\mathcal Q}$, where $U,V,W$ denotes measurements performed on different nodes, arising from positivity $\sum_\pm\mathcal Q(\pm,\pm,\mp|UVW)\ge0$. 

Again, since only two dichotomic measures are performed by each party, the maximal violation of quantum theory is attained for projective measurements and a pure $N$-qubit state. And thus those measurements, e.g., $A_x$, can be regarded as qubit observable with unit Bloch vectors so that, e.g., $(A_0C_0)^2=1$ and $(B_0+B_1)^2+(B_0-B_1)^2=4$, proving the maximal device independent upper bound for quantum violation as
$$S_N\le 2(N-2)+\sqrt 2\cdot \sqrt{\langle A_1(B_0-B_1)C_1\ldots W_1\rangle_P^2+\langle A_0(B_0+B_1)\rangle_P^2}\le 2(N-2)+\sqrt 2\cdot \sqrt{\langle (B_0-B_1)^2+(B_0+B_1)^2\rangle_P}.$$ 
This maximal violation is attained for $N$-qubit GHZ state with measurements $A_1=C_1=\ldots =W_1=X$ and $A_0=C_0=\ldots W_0=Z$ together with $B_{0,1}=\frac{Z\pm X}{\sqrt2}$. The noisy GHZ state, i.e., 
$$\rho_\eta=\eta |{\rm GHZ}_N\rangle\langle {\rm GHZ}_N|+(1-\eta)\frac I{2^N}$$ 
will also give rise to violation to our Bell-type inequality as long as $\eta>\eta_N$ where the noise threshold reads
$$\eta_N:=\frac{N-1}{N-2+\sqrt 2}$$ which is significantly better than that of Ref. \cite{royl} for $N<10$ (Eq.(\ref{nt}) below), as shown in Fig. \ref{fig:Comparison}.

\begin{figure}
	\centering
	\includegraphics[width=0.6\linewidth]{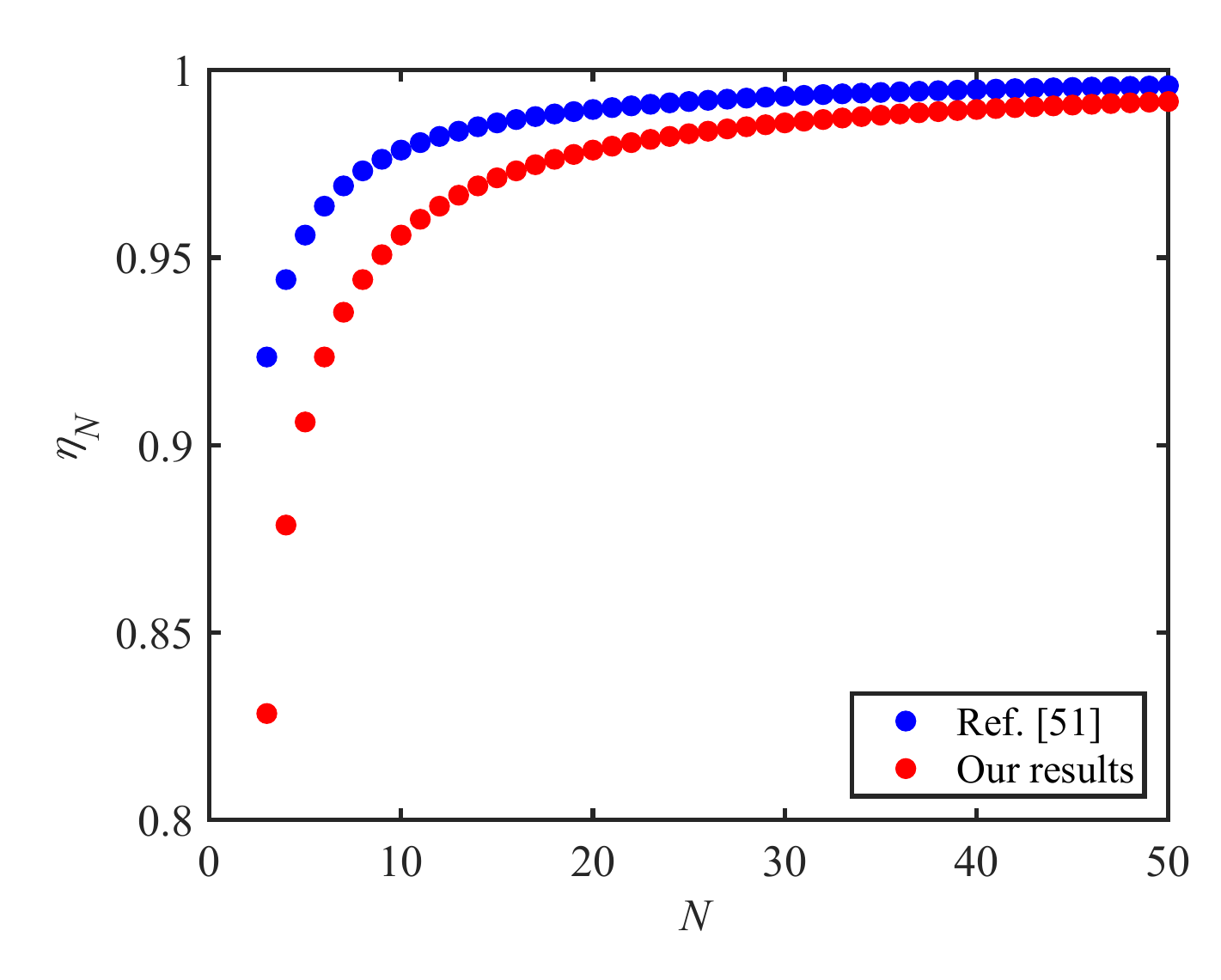}
	\caption{Comparison of white noise threshold between ours and that of Ref. \cite{royl}.
	}
	\label{fig:Comparison}
\end{figure}

\section{Estmation of fidelities of experimentally produced GHZ states $\ket{\text{GHZ}_4}$ and $\ket{\text{GHZ}_3}$}

\begin{figure}
	\centering
	\includegraphics[width=0.6\linewidth]{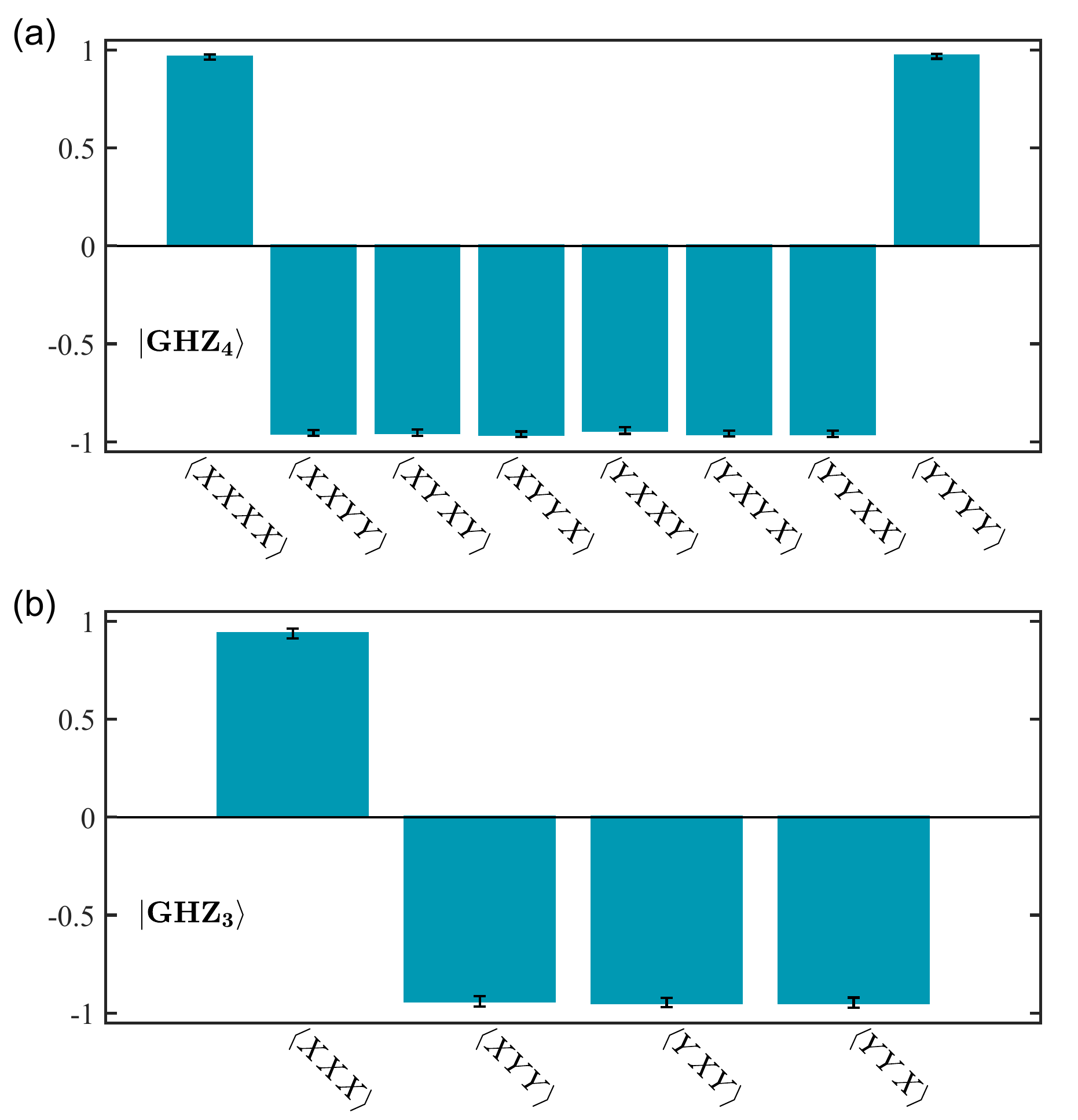}
	\caption{(a)Expectation values of experimentally generated $\ket{\text{GHZ}_4}$ measured respectively with eight bases: $XXXX$, $XXYY$, $XYXY$, $XYYX$, $YXXY$, $YXYX$, $YYXX$, $YYYY$, and $ZZZZ$.(b)Expectation values of experimentally generated $\ket{\text{GHZ}_3}$ measured respectively with four bases: $XXX$, $XYY$, $YYX$, $YXY$.}
	\label{FIGS2}
\end{figure}

We write the density matrix of ideal state $\ket{\text{GHZ}_4}$ in terms of Pauli matrices:
\begin{equation}
	\begin{aligned}
		\ket{\text{GHZ}_4}\bra{\text{GHZ}_4}=&\frac{1}{2}(\ket{HHHH}\bra{HHHH}+\ket{VVVV}\bra{VVVV})+\frac{1}{16}(XXXX-XXYY\\-&XYXY-YXXY-XYYX-YXYX-YYXX+YYYY),
	\end{aligned}
\end{equation}
where $Z=\ket{H}\bra{H}-\ket{V}\bra{V}$, $X=\ket{D}\bra{D}-\ket{A}\bra{A}$, $Y=\ket{R}\bra{R}-\ket{L}\bra{L}$, with $\ket{D}=(\ket{H}+\ket{V})/\sqrt{2}$, $\ket{A}=(\ket{H}-\ket{V})/\sqrt{2}$, $\ket{R}=(\ket{H}+i\ket{V})/\sqrt{2}$, and $\ket{L}=(\ket{H}-i\ket{V})/\sqrt{2}$.

To determine the fidelity of the generated four-photon state after post-selection, we performed measurements in nine measurement bases separately in the experiment, $XXXX$, $XXYY$, $XYXY$, $XYYX$, $YXXY$, $YXYX$, $YYXX$, $YYYY$ and $\ket{HHHH}\bra{HHHH}+\ket{VVVV}\bra{VVVV}$. The expectation value measured with base $\ket{HHHH}\bra{HHHH}+\ket{VVVV}\bra{VVVV}$ in the experiment of the main text is $0.9903\pm0.0069$ and the ones measured with other eight bases are shown in Fig.~\ref{FIGS2}(a). We then estimated the fidelity of the generated four-photon state to be $F=0.9740\pm0.0043$ with respect to the ideal state $\ket{\text{GHZ}_4}$. Accordingly, the visibility is estimated to be $\eta_4=(16F-1)/15=0.9723\pm0.0046$.

Similarly, the density matrix of $\ket{\text{GHZ}_3}$ can be decomposed as
\begin{equation}
	\begin{aligned}
		\ket{\text{GHZ}_3}\bra{\text{GHZ}_3}=\frac{1}{2}(\ket{HHH}\bra{HHH}+\ket{VVV}\bra{VVV})+\frac{1}{8}(XXX-YXY-XYY-YYX).
	\end{aligned}
\end{equation}
The expectation value measured with base $\ket{HHH}\bra{HHH}+\ket{VVV}\bra{VVV}$ in the experiment of the main text is $0.9882\pm0.0117$ and the ones measured with other four bases are shown in Fig.~\ref{FIGS2}(b). We then estimated the fidelity of the generated four-photon state to be $F=0.9653\pm0.0087$ with respect to the ideal state $\ket{\text{GHZ}_3}$. Accordingly, the visibility is estimated to be $\eta_3=(8F-1)/7=0.9603\pm0.0099$.

\section{Experimental violation of original inequalities of Ref. \cite{royl,roya} with GHZ states $\ket{\text{GHZ}_4}$ and $\ket{\text{GHZ}_3}$}

In the test for LOSR genuine multipartite nonlocality in network proposed in Ref.\cite{royl,roya}, a general network with $N$ nodes are labelled by Alice, Bob, and Charlie[$i$] ($i\in$\{$1,2,......,N-2$\}), respectively. Each party performs two alternative dichotomic measurements with outcomes $\{a,c_{[i]}\}\in\{-1,1\}$, respectively, except Bob who performs three alternative measurements $\{B_0,B_1,B_2\}$. The device-independent test is given by 

\begin{equation}
	\begin{aligned}
		I^{\tilde{C}_1=1}_{\text{Bell}} +\frac{4I_{\text{Same}_N}}{1+\langle\tilde{C}^1_1\rangle}\leq 6+\frac{4(N-2)-4\langle\tilde{C}^1_1\rangle}{1+\langle\tilde{C}^1_1\rangle},
	\end{aligned}
	\label{Original EQ}
\end{equation}
where
\begin{equation}
	\begin{aligned}
		I^{\tilde{C}_1=1}_{\text{Bell}}=\langle{A_0B_0}\rangle_{\tilde{C}_1=1} +\langle{A_0B_1}\rangle_{\tilde{C}_1=1}+\langle{A_1B_0}\rangle_{\tilde{C}_1=1}-\langle{A_1B_1}\rangle_{\tilde{C}_1=1},
	\end{aligned}
\end{equation}
with $	\tilde{C}_1 = C_{1[1]} C_{1[2]}\ldots C_{1[N-2]}$ and
\begin{equation}
	\begin{aligned}
		I_{\text{Same}_N}=\langle{A_0B_2}\rangle+\langle{B_2C_{0[1]}}\rangle+\langle{C_{0[1]}C_{0[2]}}\rangle+\ldots+\langle{C_{0[N-3]}C_{0[N-2]}}\rangle.
	\end{aligned}
\end{equation}

\begin{figure}
	\centering
	\includegraphics[width=0.6\linewidth]{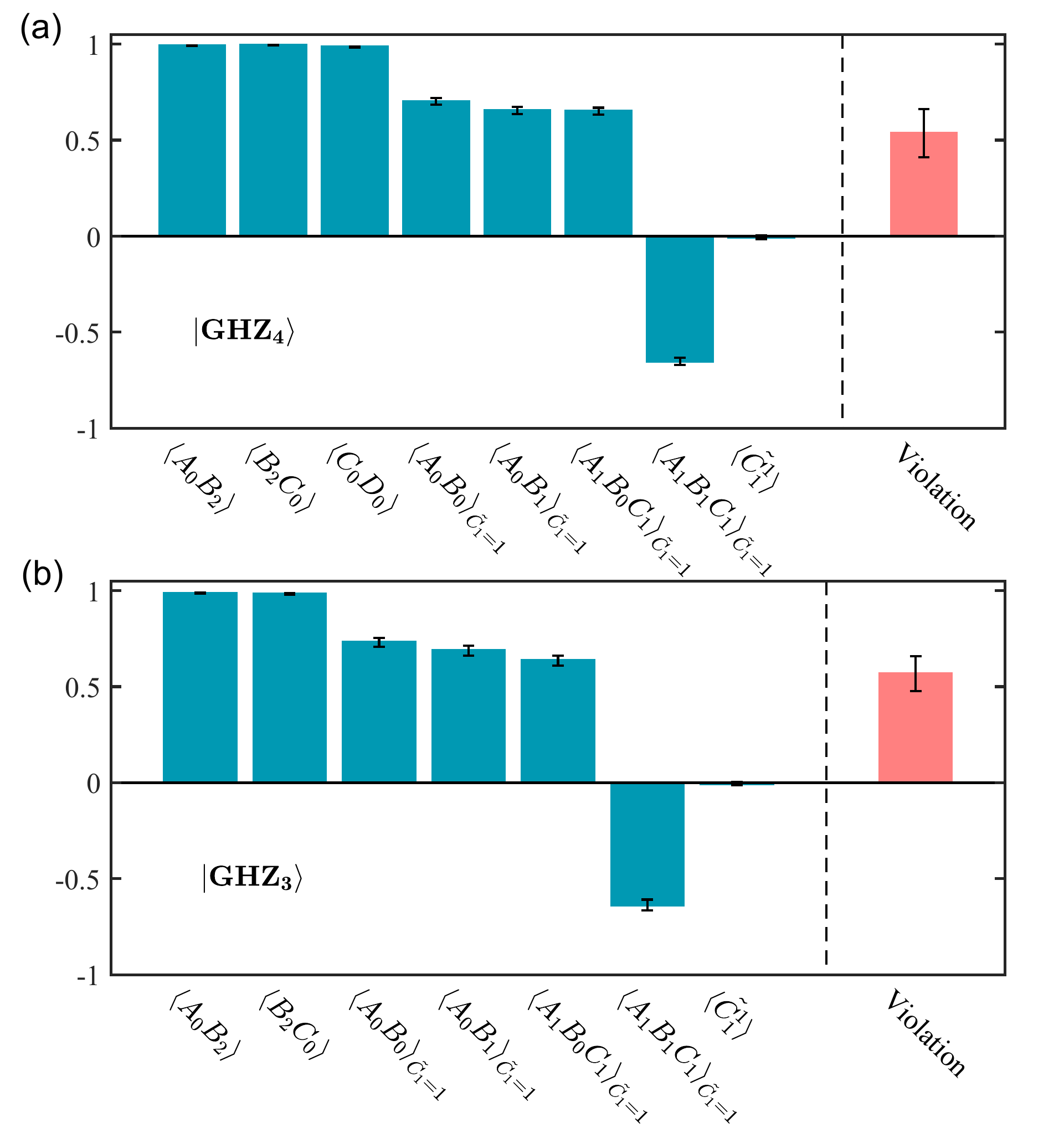}
	\caption{Experimental measurements of correlation function (left panel) and violations (right panel) of Inequality (\ref{Original EQ}) of Ref.~\cite{roya} with GHZ states $N=4$ in (a) and $N=3$ in (b). Error bars represent one standard deviations in experiments.}
	\label{FIGS3}
\end{figure}

By including a measurement setting $B_2$ in addition to the ones described in the main text, we obtain the correlations and observe the violations of Inequality (\ref{Original EQ}) (which is Inequality (14) in Ref. \cite{roya}) by $0.5361\pm 0.1258$ and $0.5685\pm0.0905$ for the states $\ket{\text{GHZ}_4}$ and $\ket{\text{GHZ}_3}$, respectively. The results are shown in Fig.~\ref{FIGS3}. Moreover the noise threshold for GHZ state in this test reads
\begin{equation}\label{nt}
	\eta_N'=\frac{2N-1}{2N-2+\sqrt 2}.
\end{equation}

\begin{thebibliography}{99}
	\bibitem{bell} J. S. Bell, On the Einstein-Podolsky-Rosen paradox, Physics 1, 195 (1964).
	\bibitem{chsh} J. F. Clauser, M. A. Horne, A. Shimony, and R. A. Holt, Proposed Experiment to Test Local Hidden-Variable Theories, Phys. Rev. Lett. 23, 880 (1969).
	\bibitem{bellnon} N. Brunner, D. Cavalcanti, S. Pironio, V. Scarani, and S. Wehner, Bell nonlocality, Rev. Mod. Phys. 86, 419 (2014).
	
	\bibitem{exp1} S. J. Freedman and J. F. Clauser, Experimental Test of Local
	Hidden-Variable Theories, Phys. Rev. Lett. 28, 938 (1972). 
	\bibitem{exp2}  A. Aspect, P. Grangier, and G. Roger, Experimental Tests of Realistic Local Theories via Bell's Theorem, Phys. Rev.
	Lett. 47, 460 (1981).
	
	
	\bibitem{cexp2} B. Hensen et al., Loophole-free Bell inequality violation
	using electron spins separated by 1.3 kilometres, Nature
	(London) 526, 682 (2015).
	\bibitem{cexp3} L. K. Shalm, E. Meyer-Scott, B. G. Christensen, P.
	Bierhorst, M. A. Wayne et al., Strong Loophole-Free Test
	of Local Realism, Phys. Rev. Lett. 115, 250402 (2015). 
	
	\bibitem{cexp4}  M. Giustina, M. A. M. Versteegh, S. Wengerowsky, J. Handsteiner, A. Hochrainer et al., Significant-Loophole-Free Test of Bell's Theorem with Entangled Photons, Phys.
	Rev. Lett. 115, 250401 (2015).
	\bibitem{cexp5} W. Rosenfeld, D. Burchardt, R. Garthoff, K. Redeker, N.
	Ortegel, M. Rau, and H. Weinfurter, Event-Ready Bell Test Using Entangled Atoms Simultaneously Closing Detection and Locality Loopholes, Phys. Rev. Lett. 119, 010402 (2017).
	
	\bibitem{cexp6}  M.-H. Li, C. Wu, Y. Zhang, W.-Z. Liu,B. Bai, Y. Liu, W. Zhang, Q. Zhao, H.Li, Z. Wang, L. You, W. J. Munro, J. Yin, J. Zhang, C.-Z. Peng, Q. Zhang, J. Fan and J.-W. Pan,  Test of Local Realism into the Past without Detection and Locality Loopholes, Phys. Rev. Lett. 8, 080404(2018).
	
	\bibitem{tel0} C. H. Bennett, G. Brassard, C. Cr\'epeau, R. Jozsa, A. Peres, and W. K. Wootters, Teleporting an Unknown Quantum State Via Dual Classical and Einstein-Podolsky-Rosen
	Channels, Phys. Rev. Lett. 70, 1895 (1993).
	
	\bibitem{qkd1} A. K. Ekert, Quantum cryptography based on Bell's theorem, Phys. Rev. Lett. 67, 661 (1991).
	
	\bibitem{qkd3} A. Ac\'in, N. Brunner, N. Gisin, S. Massar, S. Pironio, and V. Scarani, Device-independent security of quantum cryptography against
	collective attacks, Phys. Rev. Lett. 98, 230501 (2007).
	\bibitem{qkd4} N. Gisin, G. Ribordy, W. Tittel, and H. Zbinden, Quantum cryptography, Rev. Mod. Phys. 74, 145 (2002).
	
	
	\bibitem{ran1} R. Colbeck, Quantum and relativistic protocols for secure multi-party computation, Ph.D. thesis, University of Cambridge, 2011; arXiv:0911.3814
	\bibitem{ran2} R. Colbeck and A. Kent, Private randomness expansion with untrusted devices, J. Phys. A: Math. Theor. 44, 095305 (2011).
	\bibitem{ran5} A. Ac\'in and L. Masanes, Certified randomness in quantum physics, Nature (London) 540, 213 (2016).
	
	
	\bibitem{mcrp} L. Aolita, R. Gallego, A. Cabello, and A. Ac\'in, Fully Nonlocal, Monogamous, and Random Genuinely Multi- partite Quantum Correlations, Phys. Rev. Lett. 108, 100401
	(2012).
	
	\bibitem{qc1} J. I. Cirac, A. K. Ekert, S. F. Huelga, and C. Macchiavello, Distributed quantum computation over noisy channels, Phys. Rev. A 59, 4249 (1999).
	\bibitem{qc2} M. Howard and J. Vala, Nonlocality as a benchmark for universal quantum computation in Ising anyon topological quantum computers, Phys. Rev. A 85, 022304 (2012).
	\bibitem{qc3} M. Howard, J.J. Wallman, V. Veitch, and J. Emerson, Contextuality supplies the magic for quantum computation, Nature (London) 510, 351 (2014).
	
	\bibitem{sw1} C. Branciard, N. Gisin, and S. Pironio, Characterizing the Nonlocal Correlations Created via Entanglement Swapping, Phys. Rev. Lett. 104, 170401 (2010).
	\bibitem{sw2} C. Branciard, D. Rosset, N. Gisin, and S. Pironio, Bilocal versus non-bilocal correlations in entanglement swapping experiments, Phys. Rev. A 85, 032119 (2012).
	
	\bibitem{cmp1} C. Gross, T. Zibold, E. Nicklas, J. Esteve, and M. K. Oberthaler, Nonlinear atom interferometer surpasses classical precision limit, Nature (London) 464, 1165(2010). 
	\bibitem{cmp2} J. Tura, R. Augusiak, A. B. Sainz, T. V\'ertesi, M. Lewenstein, and A. Ac\'in, Detecting nonlocality in many-body
	quantum states, Science 344, 1256 (2014).
	\bibitem{cmp3} R. McConnell, H. Zhang, J.Hu, S. Cuk, and V. Vuletic, Entanglement with negative Wigner function of almost 3000 atoms heralded by one photon, Nature (London) 519, 439(2015).
	\bibitem{cmp4} J. Tura, G. De las Cuevas, R. Augusiak, M. Lewenstein, A. Ac\'in, and J. I. Cirac, Energy as a Detector of Nonlocality of Many-Body Spin Systems, Phys. Rev. X 7, 021005 (2017).
	
	
	\bibitem{qn1} D. Cavalcanti, M.L. Almeida, V. Scarani, and A. Ac?-n, Quantum networks reveal quantum nonlocality, Nat. Commun. 2, 184 (2011).
	\bibitem{qn2} N. Gisin, Q. Mei, A. Tavakoli, M.O. Renou, and N. Brunner, All entangled pure quantum states violate the
	bilocality inequality, Phys. Rev. A 96, 020304(R) (2017).
	\bibitem{qn3} I. \v Supi\'c, P. Skrzypczyk, and D. Cavalcanti, Measurement-device-independent entanglement and randomness estimation in quantum networks, Phys. Rev. A 95, 042340 (2017).
	\bibitem{qn4} A. Tavakoli, M.O. Renou, N. Gisin, and N. Brunner, Correlations in star networks: From Bell inequalities to network inequalities, New J. Phys. 19, 073003 (2017).
	\bibitem{qn5} M.-O. Renou, E. B\"aumer, S. Boreiri, N. Brunner, N. Gisin, and S. Beigi, Genuine Quantum Nonlocality in the Triangle Network, Phys. Rev. Lett. 123, 140401 (2019).
	\bibitem{qn6} N. Gisin, J.-D. Bancal, Y. Cai, P. Remy, A. Tavakoli, E. Z. Cruzeiro, S. Popescu, and N. Brunner, Constraints on nonlocality in networks from no-signaling and independ-
	ence, Nat. Commun. 11, 2378 (2020).
	\bibitem{qn7} T. Kriv\'achy, Y. Cai, D. Cavalcanti, A. Tavakoli, N. Gisin, and N. Brunner, A neural network oracle for quantum nonlocality problems in networks, npj Quantum Inf. 6, 70 (2020).
	\bibitem{qn8} J.-D. Bancal and N. Gisin, Non-Local boxes for networks, arXiv:2102.03597.
	\bibitem{qn9} S. Beigi and M.-O. Renou, Covariance decomposition as a universal limit on correlations in networks, arXiv:2103.14840.
	\bibitem{qn10} A. Tavakoli, A. Pozas-Kerstjens, M.-X. Luo, and M.-O. Renou, Bell nonlocality in networks, arXiv:2104.10700.
	
	\bibitem{qn11} H. J. Kimble, The quantum internet, Nature 453, 1023(2008).
	
	\bibitem{qn12} B. Frohlich, J. F. Dynes, M. Lucamarini, A. W. Sharpe, Z. Yuan, and A. J. Shields, A quantum access network, Nature 501, 69(2013).
	
	\bibitem{qn13} S.-K. Liao, W.-Q. Cai, J. Handsteiner, B. Liu, J. Yin, L. Zhang, D. Rauch, M. Fink, J.-G. Ren, W.-Y. Liu, Y. Li, Q. Shen, Y. Cao,
	F.-Z. Li, J.-F. Wang, Y.-M. Huang, L. Deng, T. Xi, L. Ma, T. Hu, L. Li, N.-L. Liu, F. Koidl, P. Wang, Y.-A. Chen, X.-B. Wang, M. Steindorfer, G. Kirchner, C.-Y. Lu, R. Shu, R. Ursin,
	T. Scheidl, C.-Z. Peng, J.-Y. Wang, A. Zeilinger, and J.-W. Pan, Satellite-relayed intercontinental quantum network, Phys. Rev. Lett. 120, 030501 (2018).
	
	\bibitem{qn14}S. Wehner, D. Elkouss, and R. Hanson, Quantum internet: A vision for the road ahead, Science 362, eaam9288(2018).
	
	\bibitem{qn15} Y.-A. Chen, Q. Zhang, T.-Y. Chen, W.-Q. Cai, S.-K. Liao, J. Zhang, K. Chen, J. Yin, J.-G. Ren, Z. Chen, S.-L. Han, Q. Yu,
	K. Liang, F. Zhou, X. Yuan, M.-S. Zhao, T.-Y. Wang, X. Jiang, L. Zhang, W.-Y. Liu, Y. Li, Q. Shen, Y. Cao, C.-Y. Lu, R. Shu,
	J.-Y. Wang, L. Li, N.-L. Liu, F. Xu, X.-B. Wang, C.-Z. Peng, and J.-W. Pan, An integrated space-to-ground quantum communication
	network over 4,600 kilometres, Nature 589, 214(2021).
	
	\bibitem{qn16} M. Pompili, S. L. N. Hermans, S. Baier, H. K. C. Beukers, P. C. Humphreys, R. N. Schouten, R. F. L. Vermeulen, M. J.
	Tiggelman, L. dos Santos Martins, B. Dirkse, S. Wehner, and R. Hanson, Realization of a multinode quantum network of
	remote solid-state qubits, Science 372, 259(2021).
	
	\bibitem{netwm3} A. Tavakoli, M.O. Renou, N. Gisin, and N. Brunner, Correlations in star networks: From Bell inequalities to
	network inequalities, New J. Phys. 19, 073003 (2017).
	\bibitem{netwm5} N. Gisin, J.-D. Bancal, Y. Cai, P. Remy, A. Tavakoli, E. Z. Cruzeiro, S. Popescu, and N. Brunner, Constraints on nonlocality in networks from no-signaling and independence, Nat. Commun. 11, 2378 (2020).
	
	\bibitem{qn17} Paolo Abiuso, Tamas Krivachy, Emanuel-Cristian Boghiu, Marc-Olivier Renou, Alejandro Pozas-Kerstjens, and Antonio Acin, Quantum networks reveal single-photon nonlocality, 
	arXiv:2108.01726V2.
	
	\bibitem{svet}
	G. Svetlichny, Distinguishing three-body from two-body non-separability by a Bell-type inequality, Phys. Rev. D 35, 3066
	(1987).
	
	\bibitem{op1} R. Gallego, L. E. W\"urflinger, A. Ac\'in, and M. Navascu\'es, Operational Framework for Nonlocality, Phys. Rev. Lett. 109, 070401 (2012).
	\bibitem{op2} J.-D. Bancal, J. Barrett, N. Gisin, and S. Pironio, Definitions of multipartite nonlocality, Phys. Rev. A 88, 014102 (2013).
	\bibitem{intrinsic} P. Contreras-Tejada, C. Palazuelos, and J.I. de Vicente, Genuine Multipartite Nonlocality Is Intrinsic to Quantum Networks, Phys. Rev. Lett. 126, 040501 (2021).
	
	\bibitem{royl} X. Coiteux-Roy, E. Wolfe, and M.-O. Renou, No Bipartite-Nonlocal Causal Theory Can Explain Nature's Correlations, Phys. Rev. Lett. 127, 200401 (2021).
	\bibitem{roya} X. Coiteux-Roy, E. Wolfe, and M.-O. Renou, companion paper, Any physical theory of nature must be boundlessly multipartite nonlocal, Phys. Rev. A 104, 052207 (2021).
	
	
	
	
	\bibitem{losr0}F. Buscemi, All entangled quantum states are nonlocal, Phys. Rev. Lett. 108, 200401(2012).
	\bibitem{losr00}D. Schmid, D. Rosset, and F. Buscemi, The type-independent resource theory of local operations and shared randomness, Quantum 4, 262 (2020).
	\bibitem{losr1}D. Schmid, T. C. Fraser, R. Kunjwal, A. B. Sainz, E. Wolfe, and R.W. Spekkens, Understanding the interplay of entanglement and nonlocality: Motivating and developing a new branch of entanglement theory, arXiv:2004.09194.
	\bibitem{losr2} E. Wolfe, D. Schmid, A. B. Sainz, R. Kunjwal, and R. W. Spekkens, Quantifying Bell: The resource theory of nonclassicality of common-cause boxes, Quantum 4, 280 (2020).
	\bibitem{losr3} K. Sengupta, R. Zibakhsh, E. Chitambar, and G. Gour, Quantum Bell nonlocality is entanglement, arXiv: 2012.06918.
	
	\bibitem{ngme} M. Navascu\'es, E. Wolfe, D. Rosset, and A. Pozas-Kerstjens, Genuine Network Multipartite Entanglement, Phys. Rev. Lett. 125, 240505 (2020).
	
	\bibitem{chao} R. Chao and B. W. Reichardt, Test to separate quantum theory from non-signaling theories, arXiv:1706.02008.
	\bibitem{bier} P. Bierhorst, Ruling out bipartite nonsignaling nonlocal models for tripartite correlations, Phys. Rev. A 104, 012210 (2021).
	
	\bibitem{infl1}  E. Wolfe, R. W. Spekkens, and T. Fritz, The inflation technique for causal inference with latent variables, J. Causal Infer. 7,
	0020 (2019).
	\bibitem{infl2}  M. Navascu\'es and E. Wolfe, The inflation technique completely solves the causal compatibility problem, J. Causal Infer. 8, 70
	(2020).
	\bibitem{infl3}  E. Wolfe, A. Pozas-Kerstjens, M. Grinberg, D. Rosset, A. Ac\'in, and M.Navascu\'es, Quantum Inflation: A General Approach to Quantum Causal Compatibility, Phys. Rev. X 11, 021043 (2021).
	
	
	\bibitem{qubit} L. Masnas, Extremal quantum correlations for N parties with two dichotomic observables per site, arXiv:quantum-ph/0512100.
	\bibitem{box}
	S. Pironio, J.-D. Bancal, and V. Scarani, Extremal correlations of the tripartite no-signaling polytope, J. Phys. A: Math. Theor. 44, 065303 (2011).
	
	\bibitem{Sun2019}
	Q.-C. Sun, Y.-F. Jiang, B. Bai, W. Zhang, H. Li, X. Jiang, J. Zhang, L. You, X. Chen, Z. Wang, Q. Zhang, J. Fan and J.-W. Pan, Experimental demonstration of non-bilocality with truly independent sources and strict locality constraints, Nat. Photon. 13, 687(2019).
	
	\bibitem{ZDLi} Z.-D. Li, Y.-L. Mao, M. Weilenmann, A. Tavakoli, H. Chen, L. Feng, S.-J. Yang, M.O. Renou, D. Trillo, T. P. Le, N. Gisin, A. Ac\'in, M.Navascu\'es, Z. Wang, and J. Fan, Testing Real Quantum Theory in an Optical Quantum Network, Phys. Rev. Lett. 128, 040402 (2022).
	
	\bibitem{Pan2001PRL}  J.W. Pan, M. Daniell, S. Matthew, G. Weihs, and A. Zeilinger, Experimental Demonstration of Four-Photon Entanglement and High-Fidelity Teleportation, Phys. Rev. Lett. 86, 4435 (2001).
	
	
	\bibitem{y1} Q. Chen, S. Yu, C. Zhang, C. Lai, and C. Oh, Test of genuine multipartite nonlocality without inequalities, Phys. Rev. Lett. 112, 140404
	(2014).
	\bibitem{y2} S. Yu and C. Oh, Tripartite entangled pure states are tripartite nonlocal, arXiv preprint arXiv:1306.5330 (2013).
	\bibitem{Huan2022} H. Cao, M.-O. Renou, C. Zhang, G. Mass, X. Coiteux-Roy, B.-H. Liu, Y.-F. Huang, C.-F. Li, G.-C. Guo, E. Wolfe, Experimental Demonstration that No Tripartite-Nonlocal Causal Theory Explains Nature's Correlations, arXiv preprint arXiv:2201.12754 (2022).
	
\end{thebibliography}
\end{document}